%% file: main.tex
\documentclass[sigconf,natbib=true]{acmart}

\AtBeginDocument{%
  \providecommand\BibTeX{{%
    \normalfont B\kern-0.5em{\scshape i\kern-0.25em b}\kern-0.8em\TeX}}}

\usepackage{xspace}
\usepackage{comment}
\usepackage[noabbrev,capitalise]{cleveref}
\usepackage{algorithm}
\usepackage{algorithmic}
\usepackage{multirow}
\usepackage{balance}
\usepackage{subfigure}
\usepackage{graphicx}
\usepackage{mdframed}
\usepackage{tcolorbox}
\usepackage{xcolor}
\usepackage{mathtools}
\usepackage{enumitem}

\usepackage{amsmath}
\usepackage{amsthm}

\usepackage{ragged2e}
\usepackage{colortbl}
\usepackage[table]{xcolor}

\definecolor{lightblue}{RGB}{173,216,230}
\definecolor{lightgreen}{RGB}{144,238,144}

\newcommand{\ie}{\emph{i.e., }}
\newcommand{\eg}{\emph{e.g., }}

\newcommand{\nosection}[1]{\vspace{2pt}\noindent\textbf{#1.}}

\newcommand{\ourmodel}{\textbf{Speeder}\xspace}

\begin{document}

\title{An item is worth one token in Multimodal Large Language Models-based Sequential Recommendation}

\author{Qiyong Zhong}
\affiliation{%
  \institution{Zhejiang University}
  \city{Hangzhou}
  \country{China}
}
\email{youngzhong@zju.edu.cn}

\author{Jiajie Su}
\affiliation{%
  \institution{Zhejiang University}
  \city{Hangzhou}
  \country{China}
}
\email{sujiajie@zju.edu.cn}

\author{Ming Yang}
\affiliation{%
  \institution{Zhejiang University}
  \city{Hangzhou}
  \country{China}
}
\email{yummyyummy183@gmail.com}

\author{Yunshan Ma}
\affiliation{%
  \institution{Singapore Management University}
  \country{Singapore}
}
\email{ysma@smu.edu.sg}

\author{Xiaolin Zheng}
\affiliation{%
  \institution{Zhejiang University}
  \city{Hangzhou}
  \country{China}
}
\email{xlzheng@zju.edu.cn}

\author{Chaochao Chen}
\affiliation{%
  \institution{Zhejiang University}
  \city{Hangzhou}
  \country{China}
}
\email{zjuccc@zju.edu.cn}

\begin{abstract}

Sequential recommendations (SR) predict users' future interactions based on their historical behavior.
The rise of Large Language Models (LLMs) has brought powerful generative and reasoning capabilities, significantly enhancing SR performance, 
while Multimodal LLMs (MLLMs) further extend this by introducing data like images and interactive relationships.
However, critical issues remain, i.e., 
(a) Suboptimal item representations caused by lengthy and redundant descriptions, leading to inefficiencies in both training and inference; 
(b) Modality-related cognitive bias, as LLMs are predominantly pretrained on textual data, limiting their ability to effectively integrate and utilize non-textual modalities; 
(c) Weakening sequential perception in long interaction sequences, where attention mechanisms struggle to capture earlier interactions, hindering the modeling of long-range dependencies.
To address these issues, we propose \ourmodel, an efficient MLLM-based paradigm for SR featuring three key innovations: 
1) Multimodal Representation Compression (MRC), which condenses item attributes into concise yet informative tokens, reducing redundancy and computational cost;
2) Modality-aware Progressive Optimization (MPO), enabling gradual learning of multimodal representations;
3) Sequential Position Awareness Enhancement (SPAE), improving the LLM’s capability to capture both relative and absolute sequential dependencies in long interaction sequences.
Extensive experiments on real-world datasets demonstrate the effectiveness and efficiency of \ourmodel. 
\ourmodel increases training speed to 250\% of the original while reducing inference time to 25\% on the Amazon dataset.

\end{abstract}

\begin{CCSXML}
<ccs2012>
 <concept>
  <concept_id>10010520.10010553.10010562</concept_id>
  <concept_desc>Computer systems organization~Embedded systems</concept_desc>
  <concept_significance>500</concept_significance>
 </concept>
 <concept>
  <concept_id>10010520.10010575.10010755</concept_id>
  <concept_desc>Computer systems organization~Redundancy</concept_desc>
  <concept_significance>300</concept_significance>
 </concept>
 <concept>
  <concept_id>10010520.10010553.10010554</concept_id>
  <concept_desc>Computer systems organization~Robotics</concept_desc>
  <concept_significance>100</concept_significance>
 </concept>
 <concept>
  <concept_id>10003033.10003083.10003095</concept_id>
  <concept_desc>Networks~Network reliability</concept_desc>
  <concept_significance>100</concept_significance>
 </concept>
</ccs2012>
\end{CCSXML}

\ccsdesc[500]{Information systems~Recommender systems}
\ccsdesc[500]{Information systems~Multimedia and multimodal retrieval}

\keywords{Sequential Recommender, Large Language Model, Multimodality}

\maketitle

\setlength{\floatsep}{4pt plus 4pt minus 1pt}
\setlength{\textfloatsep}{4pt plus 2pt minus 2pt}
\setlength{\intextsep}{4pt plus 2pt minus 2pt}
\setlength{\dbltextfloatsep}{3pt plus 2pt minus 1pt}
\setlength{\dblfloatsep}{3pt plus 2pt minus 1pt}
\setlength{\abovecaptionskip}{3pt}
\setlength{\belowcaptionskip}{2pt}
\setlength{\abovedisplayskip}{2pt plus 1pt minus 1pt}
\setlength{\belowdisplayskip}{2pt plus 1pt minus 1pt}

\input{./chapter/introduction.tex}

\input{./chapter/related.tex}
\input{./chapter/model.tex}

\input{./chapter/experiment.tex}

\section{Conclusion}
In this paper, we proposed \ourmodel, a remarkably efficient paradigm to multimodal large language models for sequential recommendation.
\ourmodel introduces 3 key components: 
(1) Multimodal Representation Compression (MRC), which efficiently reduces redundancy in item descriptions; 
(2) Sequential Position Awareness Enhancement (SPAE), which strengthens the model's ability to capture complex sequential dependencies; 
(3) Modality-aware Progressive Optimization (MPO), which progressively integrates different modalities to improve the model's understanding and reduce cognitive biases.
Through extensive experiments, \ourmodel demonstrates superior performance over baselines in terms of \textbf{VHR@1} and computational efficiency. 
Specifically, \ourmodel achieved 250\% of the training speed and 400\% of the inference speed compared to the state-of-the-art MLLM-based SR models.
Future work could focus on incorporating real-time feedback from real-world systems.

\bibliographystyle{ACM-Reference-Format}
\balance
\bibliography{sample-base}

\clearpage
\appendix
\input{./chapter/appendix.tex}

\end{document}

%% file: chapter/introduction.tex
\vspace{-0.08 in}
\section{Introduction}

Sequential recommendation (SR) \cite{boka2024survey,wang2018attention,su2023personalized} is widely used to predict users' interests based on historical interactions.
Early methods primarily follow an ID-based paradigm \cite{rendle2010factorizing,tan2016improved,wang2020global}, solely focusing on extracting item co-occurrence relations. 
To overcome the data sparsity and cold-start problems \cite{cold-start-sess,chen2023knowledge},
content-based approaches \cite{lai2022attribute,jin2023dual,liu2024enhancing} integrate auxiliary information, \eg text \cite{hou2022unisrec}, images \cite{hu2023adaptive}, and other modalities \cite{zhang2022dynamic}, into SR for enriching item representations with semantic features.
However, their performance is still constrained by low-quality multimodal representations generated from suboptimal pre-trained models \cite{221,222,223}.
\begin{figure*}[t]
\centering
\includegraphics[width=1.0\linewidth]{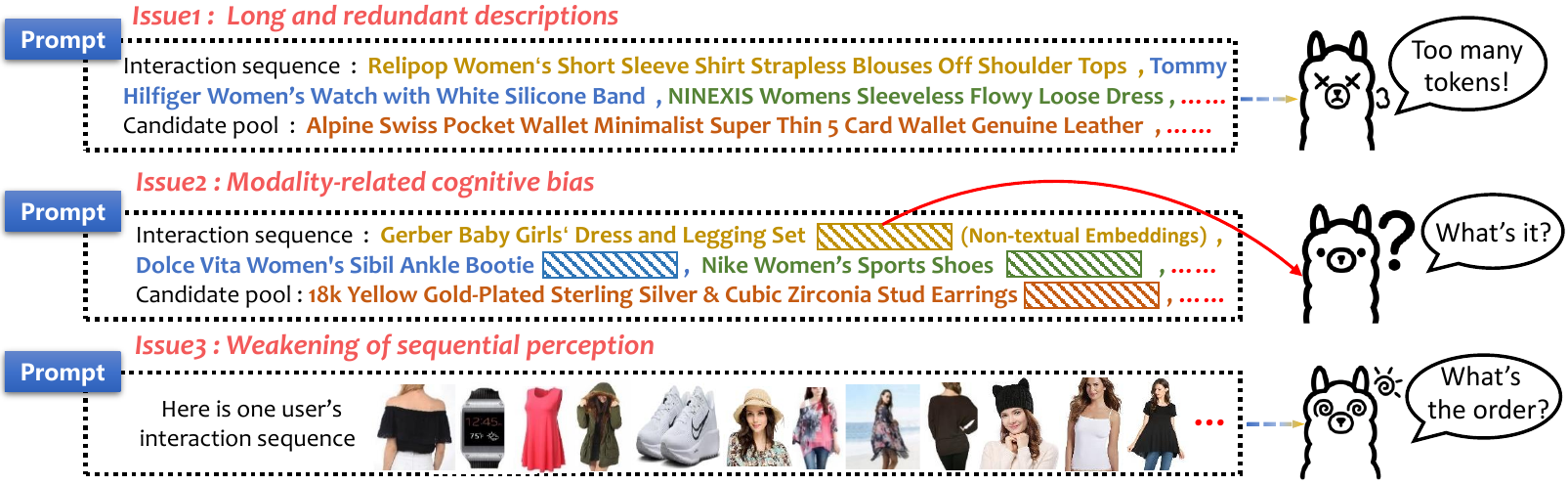}
\vspace{-0.18 in}
\caption{Motivation of \ourmodel. Issue 1 reveals suboptimal item representaions in LLM, Issue 2 implies the modality-related cognitive bias, and Issue 3 illustrates the weakening of sequential position awareness facing long sequence.
Different colors in the prompt represent distinct items, while non-textual embeddings are generated from an encoder using non-textual data.
}
\label{motivation}
\vspace{-0.06 in}
\end{figure*}

Recent advances in Large Language Models (LLMs) have demonstrated their potential in SR applications \cite{224,225,226}, driven by their powerful generative and reasoning abilities \cite{227}.
Currently, LLM-based recommendation systems can be categorized into two types: \textit{LLM-powered} and \textit{LLM-driven} models \cite{222,224,223}.
\textit{LLM-powered methods} \cite{zhang2024notellm,221} employ LLMs as advanced embedding tools to encode item content, producing rich representations to be utilized by downstream recommenders.
However, these approaches treat LLMs merely as feature extractors, overlooking their generative capabilities \cite{222,223}.
Moreover, the mismatch between LLM-generated representations and downstream embeddings prone to causing a loss of crucial semantic information \cite{liu2024large,228}, significantly degrading the performance of recommenders \cite{224,225}.
In contrast, \textit{LLM-driven methods} \cite{ye2024harnessing,226} leverage the semantic understanding and reasoning expertise of LLMs to capture user preferences, using tailored prompts to generate personalized recommendation results.

Although these efforts have significantly advanced the \textit{LLM-driven} framework, their applicability in real-world scenarios remains constrained due to the following issues as shown in~\Cref{motivation}:

\nosection{Issue1: Suboptimal item representations in LLM instructions due to long and redundant descriptions} 
(i) \textit{Inefficiency}:
Previous studies represent item titles as text \cite{llara,tmf,bunble,tallrec} in prompts, with each item consisting of multiple tokens.
But in real world, interaction sequences often contain numerous items \cite{ko2022survey,228}, the increased token count leads to low fine-tuning and inference efficiency \cite{prompt1,prompt2}. 
Several studies have attempted to address this. MLLM-MSR \cite{ye2024harnessing} breaks one long prompt into servel shorter ones, which may lead to semantic degradation and higher costs. RTA \cite{reindex} transforms multiple tokens into a single one, \eg reindexing \textit{Edge of Tomorrow} into a new token \textit{Edge-of-Tomorrow}, requiring LLM to be retrained from scratch.
(ii) \textit{Semantic redundancy on item-level}: 
Given that Natural Language Processing tasks necessitate efficient handling of various forms and out-of-vocabulary words \cite{mielke2021between,dridan2012tokenization}, current LLMs typically employ subword tokenization \cite{sennrich2015neural,kudo2018sentencepiece}.
However, the granularity of tokenization in LLMs \cite{rai2021study,yang2024rethinking} significantly differs from that required in recommendation tasks. 
Recommendation hinges on capturing the synergy between user preferences and item attributes,
with an optimal representation concisely encoding core entity characteristics \cite{ko2022survey}.
Excessive tokenization of an item introduces semantic redundancy, 
reducing the information density of individual token and blurring LLMs' cognition on core features.
(iii) \textit{Narrowed preference expression on user level}:
Item representations derived from titles \cite{tallrec,llara,tmf,bunble} reflect only a single preference aspect, 
with limited users' diverse interests expressed.
Although integrating multimodal information enhances item representations \cite{tmf,bunble}, it worsens the issue of excessive tokens in single-item descriptions, forcing researchers to make an arduous trade-off between computational cost and preference richness.
Therefore, reducing semantic redundancy in long item representations while efficiently compressing multimodal features is important.

\nosection{Issue2: Modality-related cognitive bias due to rough training strategy} 
Owing to the predominance of textual data in the pre-training phase of LLMs, 
toghether with the absence of specialized training in understanding other modalities \cite{llmsurvey},
there exists modality-related cognitive bias when LLMs deal with non-textual modality like vision and audio \cite{li2023e4srec,bunble} .
Previous studies \cite{llara,tmf} apply curriculum learning or staged training \cite{bunble} to gradually introduce multimodal data. 
However, such direct injection of multimodal signals into LLM's semantic space impairs non-textual comprehension, ultimately hindering convergence. 
This problem is exacerbated when a pre-fused multimodal representation is introduced \cite{tmf,bunble}.
Therefore, a central task in current research is effectively guiding LLMs to progressively comprehend multimodal information.

\nosection{Issue3: Weakening of sequential perception due to long interaction sequences} 
When modeling long item interaction sequences, LLMs rely on attention mechanisms \cite{228,llmsurvey},
but research \cite{hou2024large,liu2025effects,levy2024same} indicates that as sequence length grows, attention to earlier interactions decreases, impairing the model's ability to capture sequence order.
Additionally, the growing prompt length introduces more non-interactive tokens, further obscuring sequential dependencies. 
The weakened awareness of sequential order %
impedes the LLM from discriminating 
long-term and short-term user interests, resulting in worse recommendation performanc \cite{227}.
\cite{hou2024large} attempts to tackle it by proposing sequential and recency-focused prompting, which are meticulously designed prompt to direct the LLM's attention to sequential patterns. However, such naive method remains superficial and fails to enable the LLM to truly perceive the interaction sequence.
Thus, a key aspect of long-sequence SR tasks is improving the model's ability to capture sequential pattern.

To address the aforementioned issues, we follow the \textit{LLM-driven} paradigm and propose \underline{\textbf{S}}equential  
 \underline{\textbf{P}}osition-aware  \underline{\textbf{E}}nhancement with 
  \underline{\textbf{E}}fficient   
  Multimo\underline{\textbf{d}}al  
     R\underline{\textbf{e}}presentation   Comp\underline{\textbf{r}}ession (\ourmodel).
(1) To tackle the problem of suboptimal item representation in LLMs \textbf{(Issue 1)}, 
\ourmodel introduces Multimodal Representation Compression (MRC),
utilizing pre-trained encoders to compress multimodal data into compact but informative embeddings.
For improved modality integration, MRC includes Mixture of Modality Experts (MoME), which can dynamically select experts to seize modality-specific features.
(2) To overcome modality-related cognitive bias (\textbf{Issue 2)}, \ourmodel proposes Modality-wise Progressive Optimization (MPO), which adjusts model architecture together with training data in 3 stages.
MPO enables \ourmodel to gradually understand each modality and recommendation task, facilitating smoother convergence.
Furthermore, a tanh gating mechanism is incorporated to prevent the LLM from undermined by the sudden introduction of non-textual modalities.
(3) To solve weakening of sequential perception \textbf{(Issue 3)}, \ourmodel introduces Sequential Position Awareness Enhancement (SPAE) to boost LLM’s capability of perceiving the sequential patterns in interaction sequences. 
SPAE utilizes Position Proxy Task (PPT), forcing the LLM to concentrate on relative positional relationships between items in the interaction sequence, and Position Prompt Learning (PPL), which adds independently trainable embedding vectors with items to strengthen absolute position representation for every item. 
A dynamic truncation strategy is applied to handle varying sequence lengths.

The contributions are as follows:
(1) We introduce \ourmodel, a remarkably efficient \textit{MLLM-driven} SR model.
(2) The proposed MRC demonstrates the effectiveness of employing a single token for item representation in the prompt.
Even when using embeddings that are not visually interpretable to human as representations of items in the prompt, LLMs are still able to distinguish and process them.
(3) We introduce MPO to guide LLMs in progressively utilizing multimodal information.
(4) The SPAE module enhances the LLM’s ability to capture complex sequential patterns in long sequences
(5) Extensive experiments on three real-world datasets show that \ourmodel outperforms existing \textit{MLLM-based} state-of-the-art SR models.
Moreover, \ourmodel achieving 250\% of the training speed and 400\% of the inference speed against them.

%% file: chapter/model.tex
\section{METHODOLOGY}
\subsection{Problem Formulation}
We define a typical scenario as follows.
We use $\mathcal{V} = \{v_1, v_2, \ldots, v_{|\mathcal{V}|}\}$ to denote the item set in datasets.
Each interaction sequence $S_i = \{v_1, v_2, \ldots, v_n\}$ records items interacted with by the user in chronological order, where $n$ represents the length of the sequence.
Every item $v_i$ is represented by its raw modalities, \ie $v_i = \{v^{text}_i, v^{vis}_i, v^{id}_i\}$, where $v^{text}_i$ donates the textual data, $v^{vis}_i$ refers to visual content, and $v^{id}_i$ characterizes its unique identifier.
In this paper, $v^{text}$ is treated as the title of items, $v^{vis}$ as the image, and $v^{id}$ as the item ID.
Multimodal SR aims to predict the top-$K$ items from $\mathcal{V}$ that are most likely to be engaged with by the user next, based on the given interaction history. 
\ourmodel targets to guide the LLM in accomplishing the recommendation task by leveraging prompts embedded with a carefully generated multimodal item representation.

\subsection{An overview of \ourmodel}

\begin{figure*}[t]

\centering
\includegraphics[width=1.0\linewidth]{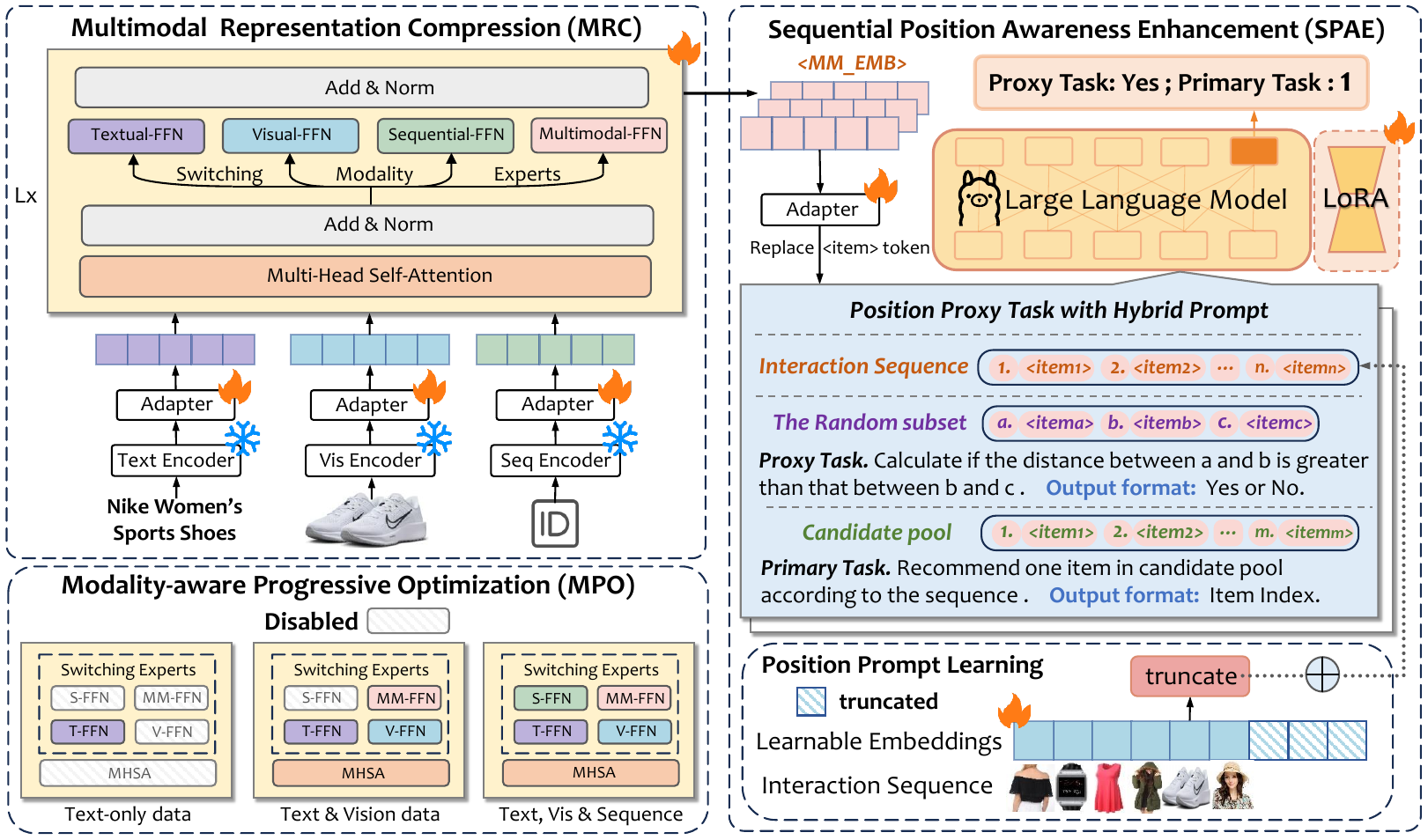}
\caption{The framework of \ourmodel. \ourmodel contains three modules. First, MRC condenses item's attributes into a compact but informative token. 
Second, SPAE augments LLM's ability of capturing the relative and absolute order in the interaction sequence.
Third, MPO allows \ourmodel to gradually understand and utilize multimodal representation. 
MHSA, T-FFN, V-FFN, S-FFN, MM-FFN are short for Multi-Head Self-Attention, Textual-FFN, Visual-FFN, Sequential-FFN, Multimodal-FFN, respectively.} 

\label{framework}
\end{figure*}

\ourmodel is presented in~\Cref{framework}, which consists of three modules: 
(i) Multimodal Representation Compression (MRC) condenses multimodal item data into dense embeddings, with Mixture of Modality Experts (MoME) integrating these to generate a compact, informative representation.
(ii) Sequential Position Awareness Enhancement (SPAE) consists of Position Proxy Task (PPT) and Position Prompt Learning (PPL). PPT directs the LLM's attention to the relative order of items in the interaction sequence, while PPL strengthens the absolute positional representations, both improving the LLM's ability to capture sequential dependencies.
(iii) MPO progressively optimizes modality-specific understanding across three stages: starting with textual data and recommendation task understanding, incorporating visual data, and finally integrating full multimodal content to improve recommendation performance.

\subsection{Multimodal Representation Compression}
To efficiently compress item representations while retaining essential semantics, we propose Multimodal Representation Compression (MRC), which (i) first utilizes pre-trained encoders to extract initial multimodal features.
(ii) Then Mixture of Modality Experts (MoME) facilitates cross-modal interaction and fusion, enhancing feature integration.
Finally, MRC yields a unified, semantically rich but compact item embedding for every item.

\nosection{Item Representation}
Firstly, we acquire heterogeneous raw data encompassing textual titles, visual content, and sequential interaction records to facilitate comprehensive item representation.
\begin{equation} \boldsymbol{p}_i^{text} = \boldsymbol{E}_{t}(\boldsymbol{v}_i^{text};\Theta_t),\boldsymbol{e}_i^{text} = {A}_{t}(\boldsymbol{p}_i^{text});
\end{equation} 
\begin{equation} \boldsymbol{p}_i^{vis} = \boldsymbol{E}_{v}(\boldsymbol{v}_i^{vis};\Theta_v),\boldsymbol{e}_i^{vis} = {A}_{v}(\boldsymbol{p}_i^{vis});
\end{equation} 
\begin{equation} \boldsymbol{p}_i^{id} = \boldsymbol{E}_{s}(\boldsymbol{v}_i^{id};\Theta_s),\boldsymbol{e}_i^{id} = {A}_{s}(\boldsymbol{p}_i^{id}).
\end{equation}
Let $\boldsymbol{E}_{t}$, $\boldsymbol{E}_{v}$, and $\boldsymbol{E}_{s}$ represent the encoders pre-trained on textual, visual, and sequential modalities, respectively, 
with corresponding parameters $\Theta_t$, $\Theta_v$, and $\Theta_s$. The outputs $\boldsymbol{p}_i^{text}$, $\boldsymbol{p}_i^{vis}$, and $\boldsymbol{p}_i^{id}$ are the last hidden embeddings capturing modality-specific features.
To bridge the semantic gap across modalities, we introduce modality-specific adapters, \ie ${A}_t$, ${A}_v$, and ${A}_s$.
Owing to the strong representational power of modern pre-trained models, we utilize LLaMA-2-7B \cite{touvron2023llama} as textual encoder, BLIP-2 \cite{blip2} for visual feature extraction, and SASRec \cite{kang2018self} for modeling sequential dependencies.

\nosection{Mixture of Modality Experts}
To further refine multimodal integration, we propose Mixture of Modality Experts (MoME) inspired by \cite{vlmo}.
Given an item's multimodal embedding, MoME generates a unified representation through multi-layered modality interactions, which exhibits formidable integration abilities, capturing complementary and higher-order semantic correlations across modalities.

Specifically, we begin by stacking the multimodal embeddings for each item as follows:
\begin{equation} \boldsymbol{e}_i = ( \boldsymbol{e}_i^{text}, \boldsymbol{e}_i^{vis}, \boldsymbol{e}_i^{id} ). \end{equation}

A shared cross-modal Multi-Head Self-Attention (MHSA) mechanism aggregates information across multiple attention heads, aligning and fusing modality-specific data. For each layer $l$, we define:
\begin{equation} 
\!\!\!\!\! {MHSA}_l \!=\!{MHSA}({Q},\!{K},\!{V}) \\
\!=\!\text{Concat}(head_1, \!head_2, \dots,\! head_h)\boldsymbol{W}_o \ ,
\end{equation} 
\begin{equation} 
\begin{gathered}
where \ \ {head}_i = {Attention} (Q\boldsymbol{W}_i^Q, K\boldsymbol{W}_i^K, V\boldsymbol{W}_i^V ) \\
= \text{softmax} \left( \frac{ H_{l-1}\boldsymbol{W}_Q\boldsymbol{W}_i^Q \times (H_{l-1}\boldsymbol{W}_K\boldsymbol{W}_i^K    )^T }{\sqrt{d_h}} \right) H_{l-1}\boldsymbol{W}_V\boldsymbol{W}_i^V ,
\end{gathered}
\end{equation}
where $\boldsymbol{W}_o \in \mathbb{R}^{d_f \times d_f}$ is the output projection matrix, and $\boldsymbol{W}_Q$, $\boldsymbol{W}_K$, $\boldsymbol{W}_V \in \mathbb{R}^{d_f \times d_f}$ are the query, key, and value matrices. The learnable matrices $\boldsymbol{W}_i^Q$, $\boldsymbol{W}_i^K$, $\boldsymbol{W}_i^V \in \mathbb{R}^{d_f \times d_h}$ project the input embeddings into the attention space for each head. $d_h = {d_f}/{h}$, where $h$ is the number of heads, and $H_{l-1}$ is the output from the previous MoME block, with $e_i$ as the initial input.

MoME replaces the standard feed-forward network (FFN) in the transformer \cite{vaswani2017attention} structure  with modality-specific expert networks: textual, visual, sequential, and multimodal experts. 
Each expert consists of two layers with nonlinear transformations and activations.
For each layer $l$, MoME dynamically selects the relevant expert based on the output of the previous layer, enabling modality-specific feature extraction and enhancing processing efficiency.
The switching modality expert machanism is formulated as:
\begin{equation} {H}_l^{\prime} = \text{LayerNorm}({MHSA_l} + H_{l-1}) ,\end{equation}
\begin{equation} \boldsymbol{\text{Modality-FFN}}(H_l^{\prime}) = \text{ReLU}(H_l^{\prime}\boldsymbol{W}_1^{m_i} + \boldsymbol{b}_1^{m_i})\boldsymbol{W}_2^{m_i} + \boldsymbol{b}_2^{m_i} ,\end{equation}
\begin{equation} {H_l} = \text{LayerNorm}(\boldsymbol{\text{Modality-FFN}}(H_l^{\prime})) + H_l^{\prime} ,\end{equation}
here, 
$\text{Modality-FFN}(H_l^{\prime})$ represents the output of the modality FFN. $\boldsymbol{W}_1^{m_i}$ and $\boldsymbol{W}_2^{m_i}$ are the learnable matrices, while $\boldsymbol{b}_1^{m_i}$ and $\boldsymbol{b}_2^{m_i}$ are the associated bias.  With $m_i \in \{\text{textual, visual, sequential, multimodal}\}$, denoting the corresponding modality expert.
$H_l$ represents the final output for layer $l$  by applying a residual connection and LayerNorm.

In the previous $L_1$ layer of the MoME block, modality-specific embeddings undergo shallow inter-modality interactions, followed by independent processing within each expert. These embeddings are then passed to the later $L_2$ layer.
In contrast to earlier $L_1$ layers, the $L_2$ layer exclusively uses the multimodal expert and MHSA to capture higher-order cross-modal interactions.
Unlike conventional mixture of expert networks \cite{yang2024xmoe,huang2024harder} which apply a soft-routing strategy, our approach employs a hard-routing strategy inspired by \cite{vlmo}, where each expert processes only the representations specific to its modality, preventing cross-modality interference and ensuring focus on domain-specific tasks.

After processing the multimodal input $\boldsymbol{e}_i$ through $L = L_1 + L_2$ layers of MoME, the final multimodal representation of item $i$ is obtained directly through an adapter as follows:
\begin{equation}
\boldsymbol{e}_i^{mm} = \boldsymbol{A}_f (\text{AvgPooling}({MoME}(\boldsymbol{e}_i))) ,\end{equation}
where $\boldsymbol{e}_i^{mm}$ is the unified multimodal representation of item $i$, derived from the average-pooled output of MoME, adapted to the LLM's semantic space via the feedforward adapter $\boldsymbol{A}_f$.
Each interaction sequence and candidate pool can be expressed as follows: 
\begin{equation} { S_i^{mm}}={\{\boldsymbol{e}_1^{mm},\boldsymbol{e}_2^{mm},\ldots,\boldsymbol{e}_n^{mm}\}} ,\end{equation}
\begin{equation} { C_i^{mm}}={\{\boldsymbol{e}_1^{mm},\boldsymbol{e}_2^{mm},\ldots,\boldsymbol{e}_m^{mm}\}} ,\end{equation}
where $S_{i}^{mm}$ is the $i$-th interaction sequence and $C_{i}^{mm}$ denotes the $i$-th candidate pool,
with every item represented by $\boldsymbol{e}_{i}^{mm}$.
\subsection{\!\!\!Sequential Position Awareness Enhancement}
We propose Sequential Position Awareness Enhancement (SPAE) to improve LLM's ability of modeling complex sequential dependencies. SPAE achieves this by two mechanisms: (i) Position Proxy Task (PPT) emphasizes the relative positions of items, enhancing the LLM's awareness of \textit{item relative order}, and (ii) Position Prompt Learning (PPL) injects absolute position information, strengthening LLM's understanding of \textit{item absolute positions} within the sequence.

\noindent \textbf{Position Proxy Task (PPT).}
In the Position Proxy Task, for a given interaction sequence $S_i^{mm}$, a random subset of three item embeddings is selected from the interaction sequence as follows: 
\begin{equation} {R}_i^{mm} = {\{ \boldsymbol{e}_{a}^{mm}, \boldsymbol{e}_{b}^{mm}, \boldsymbol{e}_c^{mm} \}}, \quad {R}_i \subset {S_i^{mm}} ,\end{equation}
here, ${R}_i^{mm}$ represents a subset of 3 distinct items, randomly selected from $S_i^{mm}$. Each item is represented by its multimodal embedding $\boldsymbol{e}_i^{mm}$, with index $a$, $b$, and $c$ denoting their positions in the subset. 
To ensure data integrity, we enforce a constraint that each interaction sequence must contain at least 3 distinct items without duplicates. Sequences failing to meet this condition are discarded.

Building on this constraint, we design a specific prompt to guide the LLM in performing the proxy task. 
The label for the PPT, denoted as ${label_{PPT}}$, is assigned according to the following criteria:
\begin{equation}
{label_{PPT}}= 
\begin{cases} 
\text{yes},\!\!\!\!& \text{if } | \text{index}_a -\text{index}_b | \leq | \text{index}_b - \text{index}_c | \\
\text{no},& \text{if } | \text{index}_a -\text{index}_b | > | \text{index}_b - \text{index}_c |
\end{cases},
\end{equation}
where $index_a, index_b, index_c$ correspond to the positions of the selected items within the random subset.

To perform this task, LLM must identify item positions in the interaction sequence and compute the relative proximity of 3 items in the random subset, promoting focus on relative order and enhancing LLM's ability to capture complex sequential dependencies.

\nosection{Position Prompt Learning (PPL)}
To strengthen absolute position representation in sequences, we develop the position prompt learning.
Clearly, PPL incorporates a set of learnable embeddings that capture the absolute position pattern of items within a sequence.
These embeddings are formulated as:
\begin{equation} {PPL} = {[\boldsymbol{p}_1][\boldsymbol{p}_2][\dots][\boldsymbol{p}_i][\dots][\boldsymbol{p}_{n_{max}}]} ,\end{equation} 
here, $[\boldsymbol{p}_i]$ denotes the $i$-th learnable positional embedding, whose dimension matches the hidden layer dimension of the LLM. Each embedding is learnable and independent, and $n_{max}$ represents the maximum sequence length of user interactions over the dataset.

To handle varying lengths, we introduce a dynamic truncation strategy adjusting $PPL$ based on each sequence's actual length. The process is formulated as follows:
\begin{equation} {PPL_i} = {truncate(PPL, n_i)} = [\boldsymbol{p}_1][\boldsymbol{p}_2][\dots][\boldsymbol{p}_{n_i}],
\end{equation}
where ${PPL_i}$ donates $PPL$ embeddings for the $i$-th interaction, and $n_i$ indicates the $i$-th sequence length. The ${truncate(\cdot)}$ function is the dynamic truncation procedure retaining only the previous $n_i$ embeddings, effectively truncating the sequence to its real length.

PPL strengthens the absolute position representations of every item by adding $PPL$ embeddings to their multimodal embeddings, which can be described as follows:
\begin{equation}
\begin{gathered}
    {S_i^{\text{inter}}} = {S_i^{\text{mm}} + \text{PPL}_i} \\
    {= \left\{ \boldsymbol{e}_1^{\text{mm}} + [\boldsymbol{p}_1], \boldsymbol{e}_2^{\text{mm}} + [\boldsymbol{p}_2], \ \ldots, \ \boldsymbol{e}_{n_i}^{\text{mm}} + [\boldsymbol{p}_{n_i}] \right\}},
\end{gathered}
\end{equation}
here, $\boldsymbol{S_i^{\text{inter}}}$ represents the integrated sequence of multimodal item embeddings, augmented with $PPL$ embedding.

\nosection{Hybrid Prompt Design}
In \textit{LLM-based} SR scenario, recommendation tasks are often driven by carefully crafted prompts. 
To equip LLM with harnessing multimodal information, we introduce the \textit{Hybrid Prompt}.
Consider the scene where the user's interaction sequence is represented as $S_i = \{v_1, v_2, \dots, v_n\}$, from which, we define a candidate set $C_i = \{c_1, c_2, \dots, c_m\}$, where $m$ denotes the size of the candidate pool. 
This set includes one item of genuine interest to the user, while the remaining items are selected through random negative sampling.
The \textit{Hybrid Prompt} is designed as shown in~\Cref{hybridPrompt}.
It is worth noting that each placeholder in the \textit{Hybrid Prompt} (\eg $[item_i]$) will be replaced by its corresponding multimodal embedding, \ie $\boldsymbol{e}_i^{mm}$ during training. 
Precisely, we use ${S_{i}^{\text{inter}}}$ to substitute the placeholders in the interaction sequence, ${C_{i}^{\text{mm}}}$ for those in the candidate set, and ${R_{i}^{\text{mm}}}$ for the random subset.

\begin{figure}[t]
\centering
\begin{tcolorbox}
\begin{center}
\textbf{Hybrid Prompt}
\rule{1.0\textwidth}{0.4pt}
\end{center}

\textbf{\# Interaction Sequence: }1.\textcolor[rgb]{0.773, 0.353, 0.067}{<item\textsubscript{1}>}2.\textcolor[rgb]{0.773, 0.353, 0.067}{<item\textsubscript{2}>}\textcolor[rgb]{0.773, 0.353, 0.067}{...} n.\textcolor[rgb]{0.773, 0.353, 0.067}{<item\textsubscript{n}>}  

\rule{1.0\textwidth}{0.4pt}

\textbf{\# The Random Subset:} a. \textcolor[rgb]{0.439, 0.188, 0.627}{<item\textsubscript{a}>}  
 b. \textcolor[rgb]{0.439, 0.188, 0.627}{<item\textsubscript{b}>}  \textcolor[rgb]{0.439, 0.188, 0.627}{...} 
 c. \textcolor[rgb]{0.439, 0.188, 0.627}{<item\textsubscript{c}>}  

\textbf{\# Proxy Task:}
  Calculate if the distance between a and b is greater than that between b and c.  

\textbf{\# Output Format:}
 Yes or No.  

\rule{1.0\textwidth}{0.4pt}

\textbf{\# Primary Task:}
 Recommend one item in candidate pool according to the interaction sequence.  

\textbf{\# Candidate Pool:}
1. \textcolor[rgb]{0.329, 0.510, 0.208}{<item\textsubscript{1}>}  
 2. \textcolor[rgb]{0.329, 0.510, 0.208}{<item\textsubscript{2}>} \textcolor[rgb]{0.329, 0.510, 0.208}{...} 
 m. \textcolor[rgb]{0.329, 0.510, 0.208}{<item\textsubscript{m}>}  

\textbf{\# Output Format:}
 Item Index.
\end{tcolorbox}

\caption{The framework of Hybrid Prompt.}
\label{hybridPrompt}
\end{figure}
\begin{figure*}[t]
\centering
\includegraphics[width=1.0\linewidth]{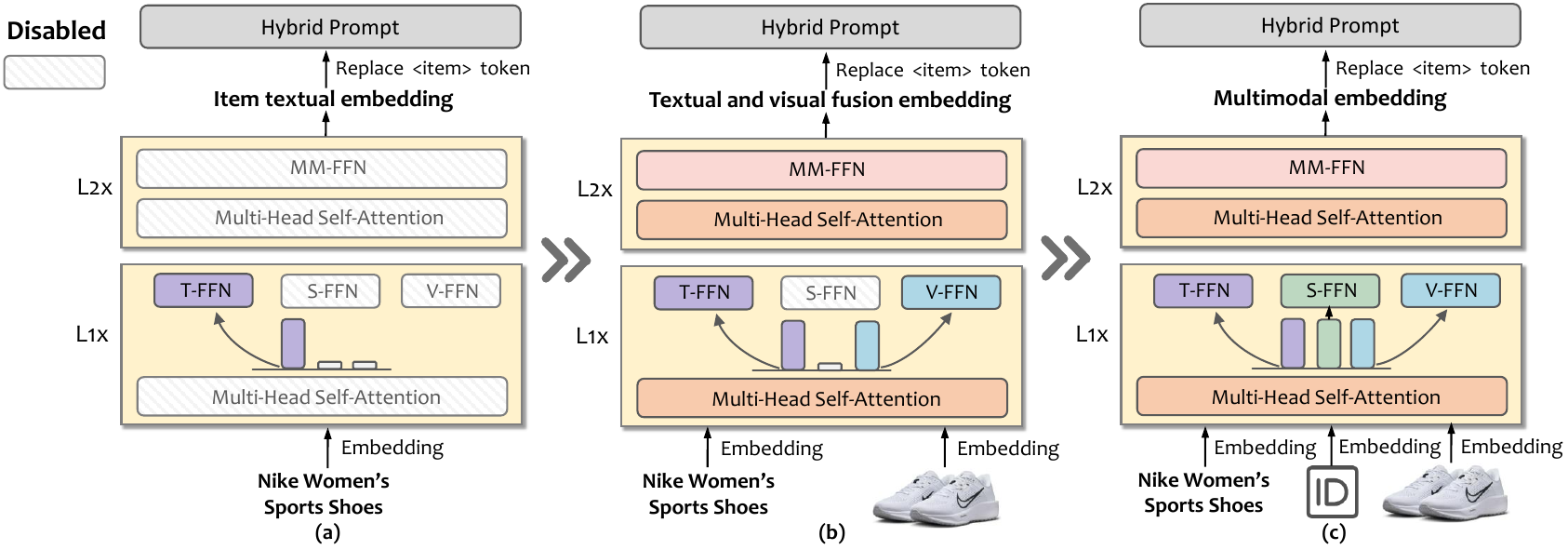}
\vspace{-0.15 in}
\caption{The pipeline of MPO. (a) In the first stage, only the Textual-FFN is trainable, using text-only data. (b) In the second stage, solely the Sequential-FFN is frozen, with training conducted on paired text and vision data. (c) In the third stage, all parameters are jointly optimized on multimodal data. Trainable parameters in the same block at different stages are shared.
} 
\label{mpo}
\end{figure*}
\subsection{Modality-aware Progressive Optimization}
To mitigate the modality-related cognitive bias arising from coarse training strategies, we propose Modality-aware Progressive Optimization (MPO) as illustrated in~\Cref{mpo}, designed to gradually guide LLMs in understanding and utilizing multimodal information. There are 3 stages in MPO, \ie 1) compressed text and task comprehension stage, 2) preliminary multimodality understanding stage, and 3) unified multimodality optimization stage.

\nosection{Stage 1: Compressed Text and Task Comprehension Stage}
In the first stage of MPO, solely textual expert is set as trainable, with the training data restricted to text-only inputs. The objective is to preserve and enhance the LLM's semantic comprehension, enabling it to accurately reconstruct item attributes from compressed textual representations while initiating learning of the SR task.

\nosection{Stage 2: Preliminary Multimodality Understanding Stage}
In the second stage of MPO, only the sequential expert is frozen, with MoME training on text-vision pairs.
To prevent the abrupt introduction of visual information from disrupting the LLM's semantic space, we apply a tanh gating mechanism inspired by \cite{alayrac2022flamingo}, to regulate non-textual features. This allows gradual assimilation of non-textual modalities while preserving learned textual priors. 
\begin{equation} \boldsymbol{e}_i^{nt} = {\text{tanh}({A}^{tan}(\boldsymbol{e}_i^{nt})) \cdot \boldsymbol{e}_i^{nt}} ,\end{equation}
where ${\boldsymbol{e}_i^{nt}}$ donates the embeddings derived from non-textual ($nt$ is short for non-textual) signals. The $\tanh$ function controls the flow of non-textual data, while $A^{tan}$ is short for $Adapter^{tan}$, which refers to a single-layer neural network adapter.

\nosection{Stage 3: Unified Multimodality Optimization Stage}
In the final stage of MPO, all parameters in MoME are jointly optimized on multimodal data, \ie textual, visual and sequential. This phase aims to fully leverage the LLM's multimodal understanding while enhancing its comprehension of SR tasks. As a result, LLM becomes more adept at capturing multifaceted preference expression on user level, improving the overall recommendation results.

\nosection{Fine-tuning with LoRA}
To reduce computational costs, we employ Parameter Efficient Fine-Tuning (PEFT) strategies, specifically utilizing the Low-Rank Adaptation (LoRA) \cite{hu2021lora}. LoRA freezes the LLM’s pre-trained weights and introduces trainable low-rank matrices for task-specific adaptations, without modifying core parameters.
We reorganize the pairwise data $\mathcal{D} = {(x_i^{s_j}, y_i^{s_j})}\ \ ,{i=1, \ldots, |\mathcal{V}|}$, 
where $x_{i}^{s_j}$ are input instructions and $y_i^{s_j}$ are true answers at stage $j$ of MoME, with each item replaced by its multimodal embedding.

We focus solely on the trainable parameters for simplicity. The loss function is therefore reformulated as follows:
\begin{equation} \label{loss1} -\sum_{\tau=1}^{\left| y^{s_j} \right|} \log p\left(y_{\tau}^{s_j} \mid x^{s_j}, y_{<\tau}^{s_j}; \Theta_{\mathrm{LoRA}}, \Theta_{\mathrm{MoME}}^{s_j}, \Theta_{\textit{PPL}}, \Theta_{\mathrm{A}}^{s_j}\right) ,\end{equation}
here, $\Theta_{\mathrm{LoRA}}$ refers to the LoRA parameters, $\Theta_{\mathrm{MoME}}^{s_j}$ represents the parameters within MoME at stage $j$, with components frozen depending on the current stage.
$\Theta_{\mathrm{PPL}}$ stands for the parameters of $PPL$, while $\Theta_{\mathrm{A}}^{s_j}$ denotes the parameters associated with all adapters. At \textit{Stage 1}, the trainable adapters are $A_t$ and $A_f$. \textit{Stage 2} extends this by incorporating $A_v$ and $A_v^{tan}$. And \textit{Stage 3} further augments the set with $A_s$ and $A_s^{tan}$, building upon the adapters from \textit{Stage 2}.

%% file: chapter/experiment.tex
\section{Experiments and analysis}

\begin{table}[t]
\small
\setlength{\tabcolsep}{3pt} %
\centering
\caption{Dataset statistics. \textit{AvgSL}: Average Sequence Length.}
\begin{tabular}{c@{\hspace{4pt}}ccc} %
\toprule
\textbf{Datasets} & \textbf{Automotive} & \textbf{Home \& Kitchen} & \textbf{Clothing \& Shoes} \\
\midrule
Inter & 170,523 & 564,252 & 865,522 \\
Item & 6,327 & 11,131 & 8,350 \\
Seq & 15,090 & 32,355 & 41,454 \\
AvgSL & 11.30 & 17.44 & 20.88 \\

\bottomrule
\end{tabular}
\label{tab:data}
\end{table}
\begin{table*}[t]
\renewcommand\arraystretch{0.9}
\centering
\caption{Experimental results on three datasets. The best results are boldfaced and the second-best results are underlined. }
{\fontsize{6.8}{7.5}\selectfont %
\label{tab:compare}
\resizebox{0.95\linewidth}{!}{
\begin{tabular}{c|ccc|ccc|ccc}
\toprule
\multirow{2}{*}{\textbf{Datasets}}
& \multicolumn{3}{c|}{\textbf{Automotive}} 
& \multicolumn{3}{c|}{\textbf{Home \& Kitchen}} 
& \multicolumn{3}{c}{\textbf{Clothing \& Shoes}} \\
\cmidrule(lr){2-4} \cmidrule(lr){5-7} \cmidrule(lr){8-10}
& HR@1 & ValidRatio & VHR@1 & HR@1 & ValidRatio & VHR@1 & HR@1 & ValidRatio & VHR@1 \\
\midrule
\textbf{GRU4Rec} & 0.2492 &	1.0000 &	0.2492&	0.2314&	1.0000&	0.2314&	0.2256&	1.0000& 0.2256	\\
\textbf{SASRec} & 0.2643 &	1.0000 &	0.2643&	0.2400&	1.0000&	0.2400&	0.2293&	1.0000& 0.2293		\\
\textbf{BERT4Rec} & 0.2801 &	1.0000 &	0.2801&	0.2765&	1.0000&	0.2765&	0.2823&	1.0000& 0.2823	 \\

\midrule
\textbf{SASRec}\textsubscript{  +Early Fusion}& 0.2694 &	1.0000 &	0.2694&	0.2356&	1.0000&	0.2356&	0.1789&	1.0000& 0.1789		\\
\textbf{SASRec}\textsubscript{  +Late Fusion} & 0.2513 &	1.0000 &	0.2513 &	0.2489 &	1.0000 &	0.2489&	0.1648&	1.0000& 0.1648		\\
\textbf{ODMT} & 0.2911 &	1.0000 &	0.2911&	0.2876&	1.0000&	0.2876&	0.3043&	1.0000& 0.3043		\\

\midrule

\textbf{Llama2} & 0.3657 &	0.1479 &	0.0541&	0.2636&	0.1732&	0.0457&	0.3768&	0.2041& 0.0769		\\
\textbf{GPT4} & 0.4651 &	0.9068 &	0.4217 &  0.3478 & 0.8679	& 0.3019	&	0.5132&	0.8394& 0.4308		\\
\textbf{TALLRec} & 0.5872 &	1.0000 &	0.5872&	0.5781&	1.0000&	0.5781&	0.6627&	0.9987& 0.6618		\\
\textbf{LLaRA} & 0.6019 &	0.9903 &	0.5961 &	0.5891&	0.9984&	0.5882&	0.6681&	1.0000& 0.6681		\\

\midrule 
\textbf{Gemini 2.0}\textsubscript{  Flash} & 0.4569 &	0.5615 &	0.2565&	0.4592&	0.4341&	0.1993&  \underline{0.7102} &	0.1473& 0.1046		\\
\textbf{Claude 3.5}\textsubscript{  Haiku} & 0.2857 &	0.0212 & 0.0061	&	0.3306 &	0.0636&	0.0204&	0.6388&	0.0383& 0.0206		\\
\textbf{TMF} & \underline{0.6327} &	1.0000 &	\underline{0.6327}&	\underline{0.6110}&	1.0000&	\underline{0.6110}&	0.7041&	1.0000& \underline{0.7041}		\\
\midrule

\ourmodel & \textbf{0.6543} & 1.0000 &	\textbf{0.6543}&	\textbf{0.6447}&	1.0000&	\textbf{0.6447}&	\textbf{0.7464}&	1.0000& \textbf{0.7464}		\\

\bottomrule

\end{tabular}
}}
\end{table*}

\subsection{Experimental Setup}
\nosection{Datasets}
We evaluate \ourmodel on 3 diverse datasets from the \textit{Amazon} platform \cite{ni2019justifying,hou2024bridging}, \ie \textit{Automotive, Home \& Kitchen, and Clothing \& Shoes}. 
For training, user behaviors are grouped into sessions based on a one-day window, 
with the latest interacted item as the prediction target.
Items with fewer than 8 interactions and sequences with fewer than 9 items are filtered out.
For multimodal learning, items with missing or invalid text/images are excluded. 
The dataset is split into training, validation, and testing sets in a ratio of 7:2:1 .
We present the datasets statistics in~\Cref{tab:data}.

\nosection{Baseline Algorithms}
\textit{(1) ID-based SR:}
GRU4Rec \cite{hidasi2015session}. 
SASRec \cite{kang2018self}.
 BERT4Rec \cite{sun2019bert4rec}.
{\textit{(2) Multimodal SR:}}
 SASRec + EarlyFusion \cite{ji2023online} (Our Extension).
 SASRec + LateFusion \cite{ji2023online} (Our Extension).
 ODMT \cite{ji2023online}.
\textit{(3) LLM-driven SR:}
 Llama2 \cite{touvron2023llama}.
 GPT-4 \cite{achiam2023gpt}.
 TALLRec \cite{tallrec}.
 LLaRA \cite{llara}.
\textit{(4) MLLM-driven SR}
 Gemini 2.0 Flash \cite{team2023gemini}
 Claude 3.5 Haiku \cite{laverghetta2025humans}.
 TMF \cite{tmf}.
The details of baseline are shown in Appendix~\ref{app:baseline}
.

\nosection{Evaluation Protocols}
Long sequences result in excessively lengthy prompts in baselines like LLaRA, with added candidate items increasing computational costs.
To mitigate this, we randomly select four non-interacted items as the candidate set, with the right item included.
Both \ourmodel and baselines aim to identify the correct item from this set.
The primary evaluation metric is Hit Rate at 1 (\textit{HR@1}), measuring the model's accuracy in predicting the correct item. 
Additionally, we employ the valid ratio (\textit{ValidRatio}) to quantify the proportion of valid responses, \ie those within the candidate set.
For non-generative models, the valid ratio is always 1.0, as they don't follow any instructions.
Some models may have high \textit{HR@1} but fail to provide effective answers. 
To address this, we propose a new metric, Valid \textit{HR@1} (\textit{VHR@1}), where $VHR@1 = ValidRatio \times HR@1$. 
All models are evaluated with the same number of candidate items and evaluation protocol for fairness.

\nosection{Implementation Details}
For all LLM-based methods requiring fine-tuning, we choose Llama2-7B as the backbone. 
Fine-tuning is conducted on 4 A40 GPUs with a batch size of 128.
To optimize training efficiency, we implement a learning rate warm-up, starting at 1/100 of the maximum rate, followed by cosine decay throughout the training. 
The embedding dimension for the LLM is 4096, and $L_1$ and $L_2$ are both fixed at 1 to mitigate overfitting.
For \ourmodel and all baselines, hyperparameters that are less critical are optimized via grid search using validation sets.
Conventional recommendations employ Adam optimizer \cite{kingma2014adam}, with a learning rate of 0.001, embedding dimension of 64, and a batch size of 128.
To reduce randomness, all experiments are conducted 5 times with distinct random seeds, and the results are averaged to ensure robustness.
\begin{table}[t]
\renewcommand\arraystretch{1.0}
\centering
\caption{Ablation study on key components of \ourmodel.}
{\fontsize{9}{8}\selectfont %
\label{ablation}
\resizebox{1.0\linewidth}{!}{
\begin{tabular}{c|ccc|ccc}
\toprule
\multirow{2}{*}{\textbf{Variants}}
& \multicolumn{3}{c|}{\textbf{Automotive}}
& \multicolumn{3}{c}{\textbf{Clothing \& Shoes}}\\
\cmidrule(lr){2-4} \cmidrule(lr){5-7} 
 &HR@1 &ValidRatio & VHR@1 &HR@1 &ValidRatio & VHR@1 \\

\midrule 
\textbf{w/o PPT} & 0.6344	& 0.9872	&0.6263	&0.7146 &0.9983 &0.7134\\
\midrule 

\textbf{w/o PPL} & 0.6417	&0.9733	&0.6246	&0.7253 &0.9942 &0.7211\\
\midrule 

\textbf{w/o SPAE} & 0.6241	&0.9712	&0.6061	&0.7088 &0.9916 &0.7028\\
\midrule 

\textbf{w/o MPO-$\mathbf{S_1}$} & 0.2653 & 1.0000 & 0.2653 & 0.7166 & 0.4696 & 0.3365 \\
\textbf{w/o MPO-$\mathbf{S_1\&S_2}$} & 0.2498 & 1.0000 & 0.2498 & 0.6498 & 0.8641 & 0.5615 \\
\midrule 

\textbf{w/o Seq-data} 	& 0.6001	&1.0000	&0.6001	&0.6987 &1.0000 &0.6987\\
\textbf{w/o Vision-data} 	& 0.5853	&1.0000	&0.5853	&0.7049 &0.9082 &0.6402\\
\textbf{w/o Text-data} 	& 0.4543	&0.9796	&0.4451	&0.6084 &0.6543 &0.3981\\
\midrule 
\textbf{w/o tanh} 	& 0.2578	&1.0000	&0.2578	&0.6742 &0.8563 &0.5773\\
\textbf{w/o ReLU} 	& 0.6456	&1.0000	&0.6456	&0.7253 &1.0000 &0.7253\\
\midrule 

\ourmodel & \textbf{0.6543} & \textbf{1.0000}& \textbf{0.6543}& \textbf{0.7464}& \textbf{1.0000}& \textbf{0.7464}\\

\bottomrule

\end{tabular}
}}
\end{table}
\subsection{Overall Performance}

\label{overall_performance}
We evaluate \ourmodel's performance against several baselines, with results presented in~\Cref{tab:compare}. 
Key observations are as follows:
1) \textbf{Multimodal information boosts recommendation performance.}
Models incorporating multimodal content overpass most ID-based approaches, highlighting the benefit of leveraging content-aware associations.
2) \textbf{Integrating LLMs enhances recommendation performance.}
Models enhanced with LLMs (TALLRec, LLaRA, etc.) consistently surpass those without LLMs, including both ID-based and modality-integrated models.
This demonstrates that the reasoning power and vast knowledge embedded in LLMs can significantly refine recommendation results.
3) \textbf{Fine-tuning enhances LLMs' performance on recommendation tasks.}
Models without fine-tuning (Llama2, GPT4, Gemini 2.0 Flash, Claude 3.5 Haiku) exhibit lower \textit{HR@1}, \textit{ValidRatio}, and ultimately \textit{VHR@1}, in comparison to fine-tuned models. 
This suggests that fine-tuning significantly improves LLMs' capability to effectively conduct recommendation tasks. 
Notably, models fine-tuned on Llama2 (TALLRec, LLaRA, TMF, \ourmodel) outperform the base Llama2 in \textit{HR@1}, \textit{ValidRatio} and \textit{VHR@1}.
4) \textbf{\ourmodel outperforms all baselines.}
Firstly, compared to traditional SR models without LLMs, 
\ourmodel excels by fully harnessing the sophisticated semantic modeling and reasoning capabilities of LLMs.
Secondly, \ourmodel surpasses advanced LLM-driven models for several reasons:
1) Unlike Llama2, GPT-4, Gemini 2.0 Flash, and Claude 3.5 Haiku, which generate recommendations without fine-tuning, \ourmodel benefits from task-specific adaptation.
2) While TALLRec and LLaRA utilize LLMs with single modalities, \ourmodel integrates multimodal inputs to capture multifaced preference expression on user level, producing better results.
3) In contrast to TMF, which suffers performance degradation from long, redundant descriptions and extended interaction sequences,
\ourmodel enables deeper modality interactions while keep prompt concise, 
and exhibits a stronger capacity to capture complex sequential patterns in lengthy interaction sequences.
\subsection{Ablation Study}
\begin{figure}[t]
\centering
\includegraphics[width=1.0\linewidth]{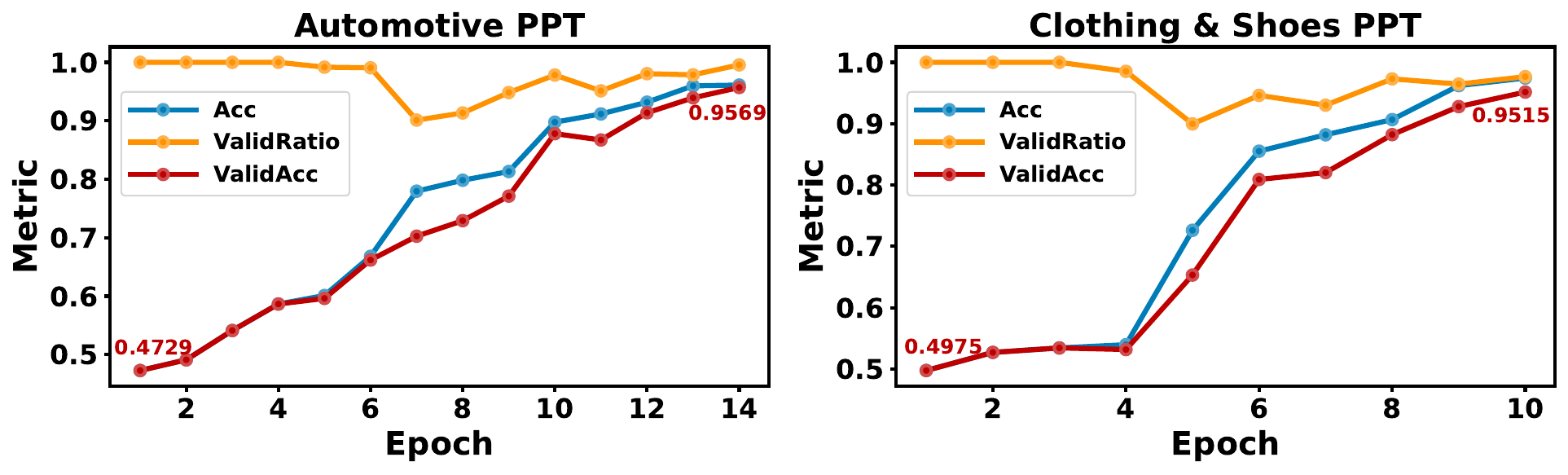}
\caption{Performance of Position Proxy Task.} 
\label{ppt}
\vspace{-0.1 in}
\end{figure}
\subsubsection{\textbf{Study of the Position Proxy Task (PPT)}} \label{ab:ppt} 
To evaluate the effectiveness of PPT, we conduct an ablation study by removing it on two datasets, denoted as \textbf{w/o PPT} in~\Cref{ablation}. 
The results indicate that removing PPT negatively impacts both \textit{HR@1} and
\textit{ValidRatio}. 
This underscores the critical role of enhancing relative position awareness for LLMs, particularly in long sequence scenarios.
\subsubsection{\textbf{Effects of Performance of Position Proxy Task}}
To validate \ourmodel's effectiveness in completing the PPT,
we present its performance in~\Cref{ppt}. The outcomes show that as the number of epochs increases,
both \textit{Acc} (Accuracy) and \textit{ValidAcc} (where \textit{ValidAcc} = \textit{Acc} $\times$ ValidRatio) steadily improve, with \textit{ValidRatio} first slightly decreases and then rises to about 1.
At the final epoch, \textit{ValidAcc} reaches approximately 95\% across two datasets, 
demonstrating \ourmodel's strong ability to perform the PPT effectively.

\subsubsection{\textbf{Study of Position Prompt Learning (PPL)}} 
To illustrate the rationality of the proposed PPL, we remove it as variant \textbf{w/o PPL} in~\Cref{ablation}.
The findings align with those in \textit{Section \ref{ab:ppt}}.
Notably, excluding PPL causes a more significant reduction in \textit{ValidRatio} compared to the removal of PPT.
\subsubsection{\textbf{Effects of Sequential Position Awareness Enhancement (SPAE)}} 
We remove SPAE as outlined in~\Cref{ablation},
the result, \ie \textbf{w/o SPAE} reveals that SPAE significantly contributes to improvements in both \textit{HR@1} and \textit{ValidRatio} for \ourmodel, 
highlighting the necessity of enhancing position awareness within interaction sequences.
\subsubsection{\textbf{Study of Modality-aware Progressive Optimization (MPO)}} 
We compare 2 variants:
(a) \textbf{w/o MPO-$\mathbf{S_1}$}, removing the first stage of MPO, initializing it with a combination of textual and visual data, 
and (b) \textbf{w/o MPO-$\mathbf{S_1 \& S_2}$}, removing both the first and second stages, sending textual, visual, and sequential data together. 
The results in~\Cref{ablation} reveal that 1) \ourmodel fails to learn anything on the \textit{Automotive} dataset without MPO,
and 2) removing MPO leads to declines in both \textit{HR@1} and \textit{ValidRatio} on the \textit{Clothing \& Shoes} dataset, with \textit{ValidRatio} decreasing more significantly.
emphasizing the importance of Modality-aware Progressive Optimization.
\subsubsection{\textbf{Effects of Specific Modalities}}
To evaluate the contributions of each modality, we test three variants, \ie
\textbf{w/o Seq-data}, \textbf{w/o Vision-data}, and \textbf{w/o Text-data} which remove sequential, visual, and textual information, respectively.
The results in~\Cref{ablation} lead to the following conclusions:
1) The absence of specific modalities affects \textit{HR@1} on the \textit{Automotive} dataset;
2) The absence of certain modalities decreases \textit{HR@1}, with \textbf{w/o Vision-data} and \textbf{w/o Text-data} both negatively affecting \textit{ValidRatio};
3) \textbf{w/o Text-data} causes the most significant damage to \ourmodel, which confirms that textual data is the most important signal to LLM.
\subsubsection{\textbf{Study of tanh gating Mechanism}}

To evaluate the impact of the proposed $tanh$ gating mechanism, we design 2 variants, \ie \textbf{w/o tanh} which removes the $tanh$ gating and \textbf{with ReLU}, replacing $tanh$ gating with $ReLu$ gating.
The results in~\Cref{ablation} show that \ourmodel with the $tanh$ gating achieves the best overall performance. 
Removing the $tanh$ gating severely disrupts training: the model fails to converge on the \textit{Automotive} dataset, and both \textit{HR@1} and \textit{ValidRatio} decrease on \textit{Clothing}. 
We also observe that, without $tanh$, \ourmodel requires twice as many epochs to converge on \textit{Clothing}. 
While using a $ReLU$ gating does not hinder convergence, its performance remains inferior to that of $tanh$.
In addition, we observed a notable phenomenon: during training,
the abrupt introduction of the visual modality data leads to a sharp drop in \textit{HR@1} within that epoch (16\% for \textit{Automotive} and 11\% for \textit{Clothing}) without $tanh$ gating,
indicating its importance to stabilize learning process.

\subsection{Training and Inference Time}
\begin{figure}[t]
\centering
\includegraphics[width=1.0\linewidth]{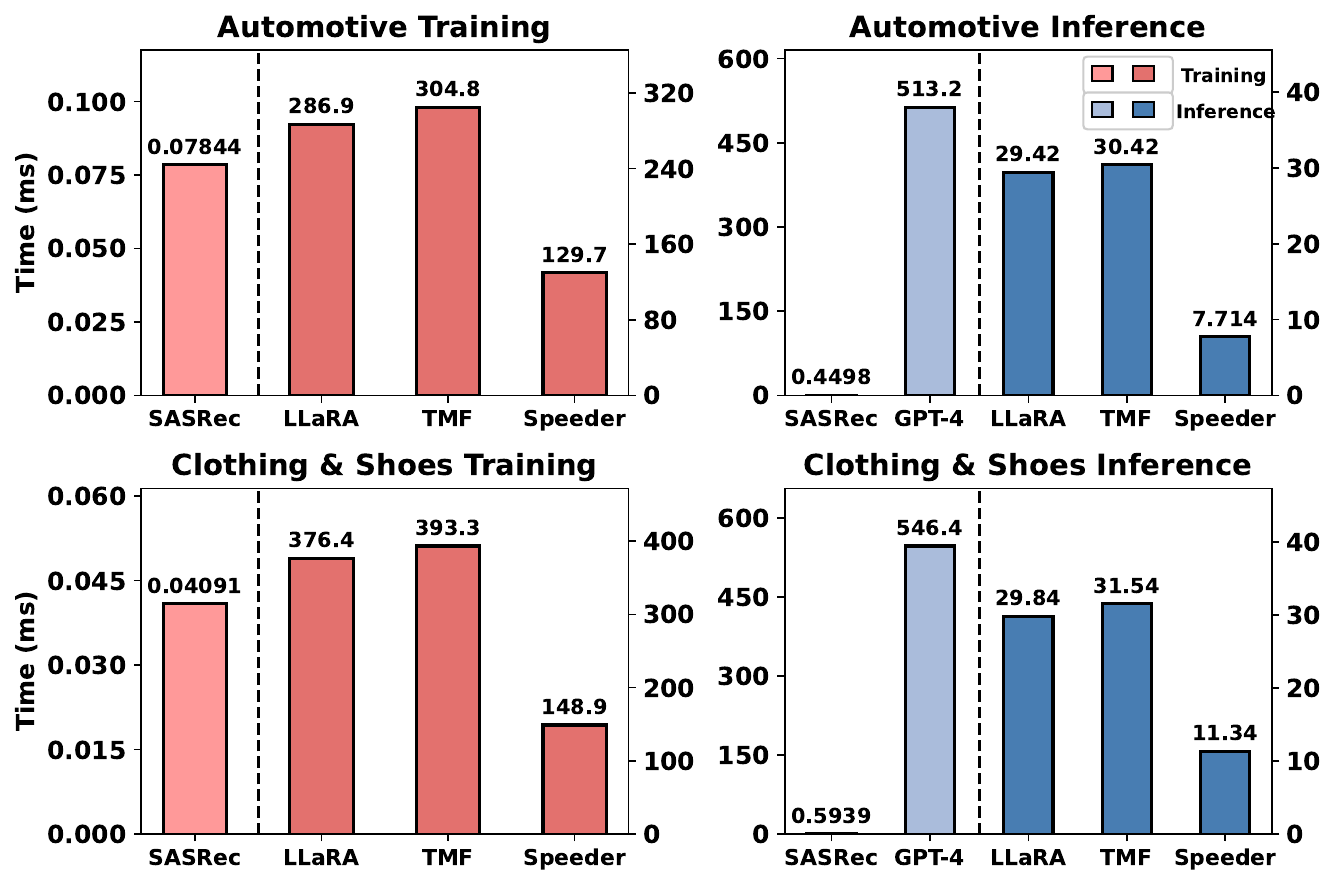}
\vspace{-0.15 in}

\caption{Time consumption of per sample.} 
\label{speed}
\vspace{-0.1 in}
\end{figure}

To evaluate the efficiency of \ourmodel during training and inference,
we compare it with SASRec (no-LLM), GPT-4 (LLM without fine-tuning on recommendation task), LLaRA (LLM-based), and TMF (MLLM-based) across two datasets as shown in~\Cref{speed}.
The \textit{y}-axis represents the average time per sample calculated under a batch size of 128, measured in milliseconds.
The findings suggest that:
1) SASRec exhibits high training and inference speeds due to its simplified model structure;
2) GPT-4 has the slowest inference speed owing to its large number of parameters (1760B); 
3) \ourmodel outperforms current state-of-the-art methods, such as LLaRA and TMF, in terms of both training and inference speed. 
Specifically, \ourmodel is 2.5 times the speed of LLaRA and TMF during training, and nearly 4 times their speed during inference,
demonstrating its efficiency as an \textit{MLLM-based} method.

\subsection{Case Study}
\begin{figure}[t]
\centering
\includegraphics[width=1.0\linewidth]{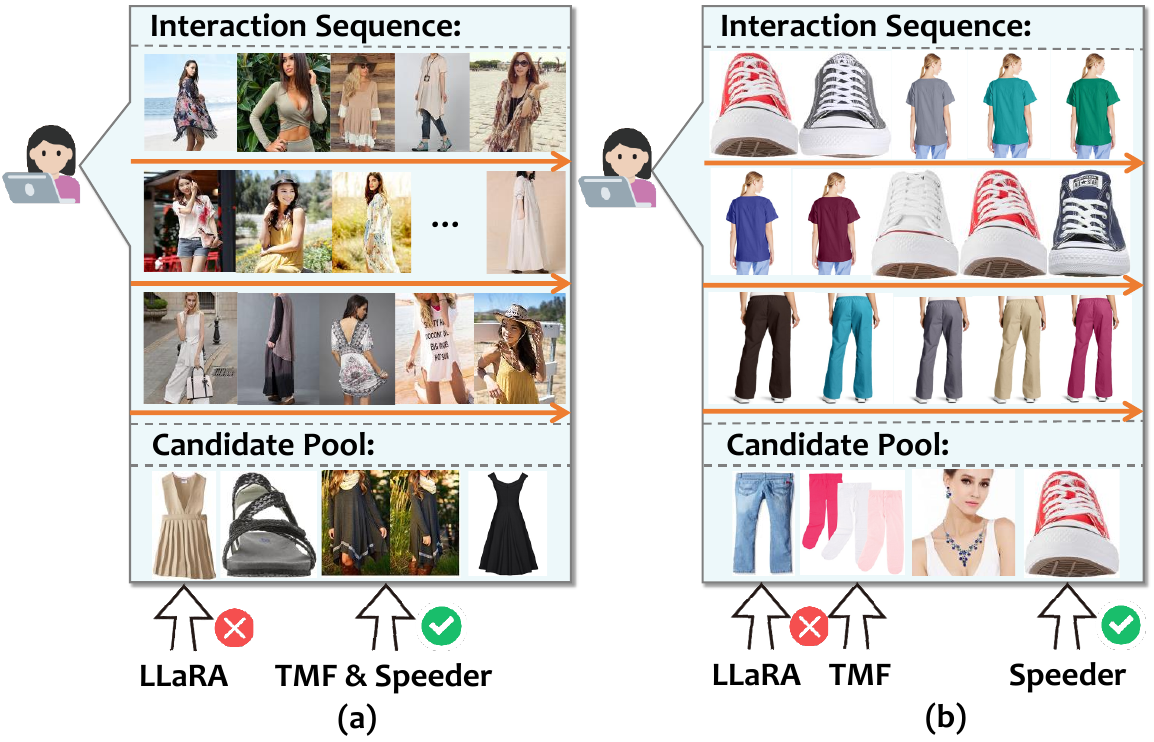}
 \vspace{-0.15 in}

\caption{Case Study. (a) The user tends to summer women's clothing worn by models in real-world scenarios, with TMF \& \ourmodel generating the right answer. (b) The user is attempting to select canvas shoes with appropriate colors that are compatible with the colors of chosen top and pants. } 
\label{casestudy}
\vspace{-0.1 in}
\end{figure}
To demonstrate the powerful capability of \ourmodel in capturing multifaceted preference expressions at the user level, 
and its superior ability to identify complex sequential patterns within interaction sequences,
we conduct a case study to compare the outputs of three models: LLaRA, TMF, and \ourmodel.
\subsubsection{\textbf{Multifaceted Preference Expression at the User Level}}
As shown in (a) of~\Cref{casestudy}, 
the user interacts with various summer women's clothing items, such as dresses, T-shirts, hot pants, and short skirts.
A common characteristic among these items is that all are worn by models in real-world scenarios, 
highlighting the user's strong preference for usage context and matching effects—preferences that are difficult to convey through non-visual information alone. 
Both TMF and \ourmodel integrate visual data, enabling them to accurately capture and reflect these nuanced preferences. In contrast, LLaRA fails to generate the correct answer.
\subsubsection{\textbf{Capturing Complex Sequential Patterns}}
As illustrated in (b) of~\Cref{casestudy}, 
the user interacts with a series of canvas shoes, tops, and casual pants. 
LLaRA selects a pair of blue jeans, assuming the user may prefer different types of pants, 
while TMF chooses casual pants in a different color from the interacted items. 
Both models fail to seize the intricate sequential patterns. 
The user is, in fact, attempting to match the color of the canvas shoes with the top and casual pants. 
Only \ourmodel successfully identifies the correct item, demonstrating its ability to capture complex sequential dependencies, 
a capability facilitated by the SPAE module.

%% file: chapter/appendix.tex
\appendix
\section{Related Work}

\nosection{LLMs for Sequential Recommendation}
Sequential Recommendation (SR) predicts the top-k items most likely to engage the user next based on their interaction history~\cite{su2023personalized,hou2022unisrec}.
Despite the wide application, they still face issues such as sparsity, cold-start, and limited representational capacity.~\cite{zheng2022ddghm,liu2023joint}.
With the powerful representational abilities \cite{llmsurvey}, LLMs hold key promise for enhancing SR performance \cite{225,223}.
Existing \textit{LLM-based} methods can be classified into two paradigms \cite{222,224,223}, \ie \textit{LLM-powered} and \textit{LLM-driven} frameworks. 
The former \cite{zhang2024notellm,luo2024molar,tian2024mmrec} synthesizes LLMs into existing systems to provide contextual or semantic enrichment \cite{221}.
However, these methods do not fully unleash generative potentials of LLMs and also suffer from significant information loss when converting LLM embeddings for downstream recommendation \cite{222,223}.
In contrast, the \textit{LLM-driven} paradigm \cite{224,225,li2023e4srec,llara} leverages designed prompts to directly guide the LLMs for recommendation. 
Unlike previous methods that rely on data with a minority modality \cite{tallrec,llara,zhang2024notellm,hu2024enhancing}, \ourmodel proposes an efficient multimodal representation compression mechanism with modality-aware progressive optimization for comprehensive LLM sequential learning.

\nosection{Multimodal Large Language Models}
Recent breakthroughs in multimodal large language models (MLLMs) \cite{mllm1,mllm2,mllm4} refine the incorporation of modality information to cohesive systems.
Current methods use technology like self-attention \cite{vilt,albef,vlmo} or cross-attention \cite{blip} for modality fusion. 
The field of recommendation systems based on MLLMs is still in its nascent stages \cite{ye2024harnessing}, with two primary approaches emerging \cite{224,225}.
The first approach \cite{zhang2024notellm-2, ye2024harnessing} integrates multimodal data into the LLM framework to enhance downstream recommendation,
while the second \cite{tmf} embeds multimodal information within the designed prompt to guide the LLM’s responses effectively .
In SR, the first paradigm struggles with utilizing LLM generative potential and embedding-related information loss \cite{223,222}. 
The second, \eg LLaRA \cite{llara}, TMF \cite{tmf}, faces high computational costs and redundancy from long item representations, leading to sequence truncation and limited sequential pattern understanding.
\ourmodel addresses these issues by compressing item representations while simultaneously enhancing the LLM's perception of interaction order.
Consequently, \ourmodel validates to be highly effective in handling multimodal long-sequence SR tasks, ensuring both computational efficiency and improved performance in capturing sequential dependencies.

\section{Baseline Algorithms}
\label{app:baseline}

\nosection{\textit{ID-based SR}}
(1) \textbf{GRU4Rec} \cite{hidasi2015session} employs gate
recurrent units to seize sequential patterns.
(2) \textbf{SASRec} \cite{kang2018self} captures sequential preference with self-attention mechanism.
(3) \textbf{BERT4Rec} \cite{sun2019bert4rec} models users' interaction preference by using a bi-directional transformer.

\nosection{\textit{Multimodal SR}}
(4) SASRec+EF \cite{ji2023online} (Our Extension) enhances SASRec by incorporating ID, text, and image inputs, with Early Fusion applied.
(5) SASRec+LF \cite{ji2023online} (Our Extension) builds upon SASRec, utilizing Late Fusion to integrate ID, text, and image features through Transformer layers.
(6) ODMT \cite{ji2023online} introduces multi-modal fusion via an ID-aware Transformer for SR.

\nosection{\textit{LLM-driven SR}}
(7) \textbf{Llama2} \cite{touvron2023llama}, an open-source large language model developed by Meta, is extensively utilized for instruction-following tasks with pre-trained parameters.
(8) \textbf{GPT-4} \cite{achiam2023gpt} is a cutting-edge LLM from OpenAI, excelling in a wide array of tasks. We obtain the results by calling the GPT-4 API, with similar procedures followed for Gemini and Claude.
(9) \textbf{TALLRec} \cite{tallrec} performs instruction tuning for LLM using recommendation datasets and tasks. The official TALLRec is limited to binary outputs (\ie yes or no). We modify it to suit our setting, where it outputs the selected item from the candidate pool.
(10) \textbf{LLaRA} \cite{llara} leverages hybrid prompts, combining text tokens with behavioral embeddings from traditional recommender systems. There are three versions of the original LLaRA, each utilizing a different sequential encoder. In our work, we adopt SASRec to implement LLaRA.

\nosection{\textit{MLLM-driven SR}}
(11) \textbf{Gemini 2.0 Flash} \cite{team2023gemini} is a fast MLLM from Google, processing audio, images, video, and text with low latency, available through API.
(12) \textbf{Claude 3.5 Haiku} \cite{laverghetta2025humans} is an MLLM developed by Anthropic, optimized for reasoning and conversation, accessible via API.
(13) \textbf{TMF} \cite{tmf} extends LLaRA by integrating visual, textual, and graph-based embeddings for items and behaviors.
since the official implementations have not been released, we follow the paper for implementation.

\section{More Analysis and Results}

\subsection{The Scalability of \ourmodel}

We conduct a comprehensive analysis of \ourmodel’s scalability from 4 perspectives under increasing data scale. In this section, a larger data scale typically refers to an expansion in the number of items, the number of sequences, and the total volume of interactions.

\nosection{Performance with Larger Data Scale}
As the item catalog expands, Multimodal Representation Compression (MRC) can learn a richer semantic space and capture deeper cross-modal interactions, thereby enhancing generalization. At the same time, longer user histories provide a more complete view of preference evolution, enabling the Sequential Position Awareness Enhancement (SPAE) to be trained and function more effectively. 
Additionally, with larger data scales, the \ourmodel architecture supports higher-dimensional modality representations by increasing the depth of $L_1$ and $L_2$ for MRC, and can also benefit from more powerful LLM backbones.
To evaluate the real-world performance of \ourmodel, we conduct experiments on three benchmark datasets from Amazon, a widely used platform in the recommendation domain. As demonstrated by the results in~\Cref{overall_performance}, \ourmodel consistently outperforms all baseline methods across these real-world datasets, achieving significant improvements.

\nosection{Space Complexity Analysis}
The total memory usage of \ourmodel is shown as follows:
\begin{equation}
O(P_{MRC} + P_{Adapters} + P_{LLM} + P_{PPL} + \mathcal{|V|} \cdot d), \tag{Eq.1}
\end{equation}
where $P_i$ is the parameter count of module $i$, $P_{PPL} = n_{\max} \cdot d$ (with $n_{\max}$ as max sequence length and $d$ as the hidden dimension of the LLM). 
We adopt a pre-cached strategy, where multimodal representations of existing items are computed in advance and stored to improve inference efficiency.
$\mathcal{|V|} \cdot d$ represents the space for storing multimodal embeddings from Multimodal Representation Compression (MRC) and Adapters, with $\mathcal{V}$ represents the item set. 
Since $P_{MRC}$, $P_{Adapters}$, and $P_{LLM}$ are constant and the rest scale linearly, \ourmodel’s storage remains manageable and supports large-scale training and deployment.

\nosection{Time Complexity Analysis}
\ourmodel’s training and inference time complexity is largely independent of the item catalog size.
As it operates in the ranking stage rather than recall, it does not perform full-item ranking.
By handling only a small candidate set via the \textit{Hybrid Prompt}, \ourmodel remains inherently decoupled from the complete item set.

But the time complexity of \ourmodel is highly based on the length of user interaction histories.
\ourmodel supports efficient caching (also named pre-cached) strategies. 
During training, embeddings obtained from frozen pre-trained encoders can be pre-cached. During inference, multimodal embeddings processed by Multimodal Representation Compression (MRC) and Adapters can also be cached.
For new items, the encoding overhead is negligible compared to LLM inference costs. Embedding lookup is $O(1)$ in time complexity.

Let $n$ denote the number of interacted items, \ie the length of user interaction histories, $m$ the size of the candidate pool, 
$t_0$ the number of non-item tokens in the \textit{Hybrid Prompt} (typically instruction tokens, with negligible variation across models and approximated as a constant),
$d$ the hidden dimension of the LLM, $T$ the number of training steps, $B$ the batch size, $L$ the number of LLM layers, and $Avg_{token}$ the average number of tokens per item.

During training, LLaRA, TMF, and \ourmodel all extract features via pre-trained encoders and Adapters. Assuming cached inputs, their computational overhead becomes:
\begin{equation}
O\left((n + m) \cdot (T_{Fusion} + T_{Adapters})\right),  \tag{Eq.2}
\end{equation}
which is minor compared to LLM. $T_i$ denotes the time cost for each item processed by module $i$. Notice that LLaRA has no Fusion module.

The main time complexity can be expressed as:
\begin{equation}
O\left(TBL \cdot [N^2 \cdot d + N \cdot d^2] \right), \tag{Eq.3}
\end{equation}
where $N = Avg_{token} \cdot (n + m) + t_0$, indicating the length of \textit{Hybrid Prompt}. 
For \ourmodel, $Avg_{token} = 2$ (only index and multimodal representation), while LLaRA and TMF reach approximately 20 in our paper due to full-text titles. 
Thus, for larger $n$, Speeder’s overhead remains significantly lower.

\begin{figure}[t]
\centering
\includegraphics[width=1.0\linewidth]{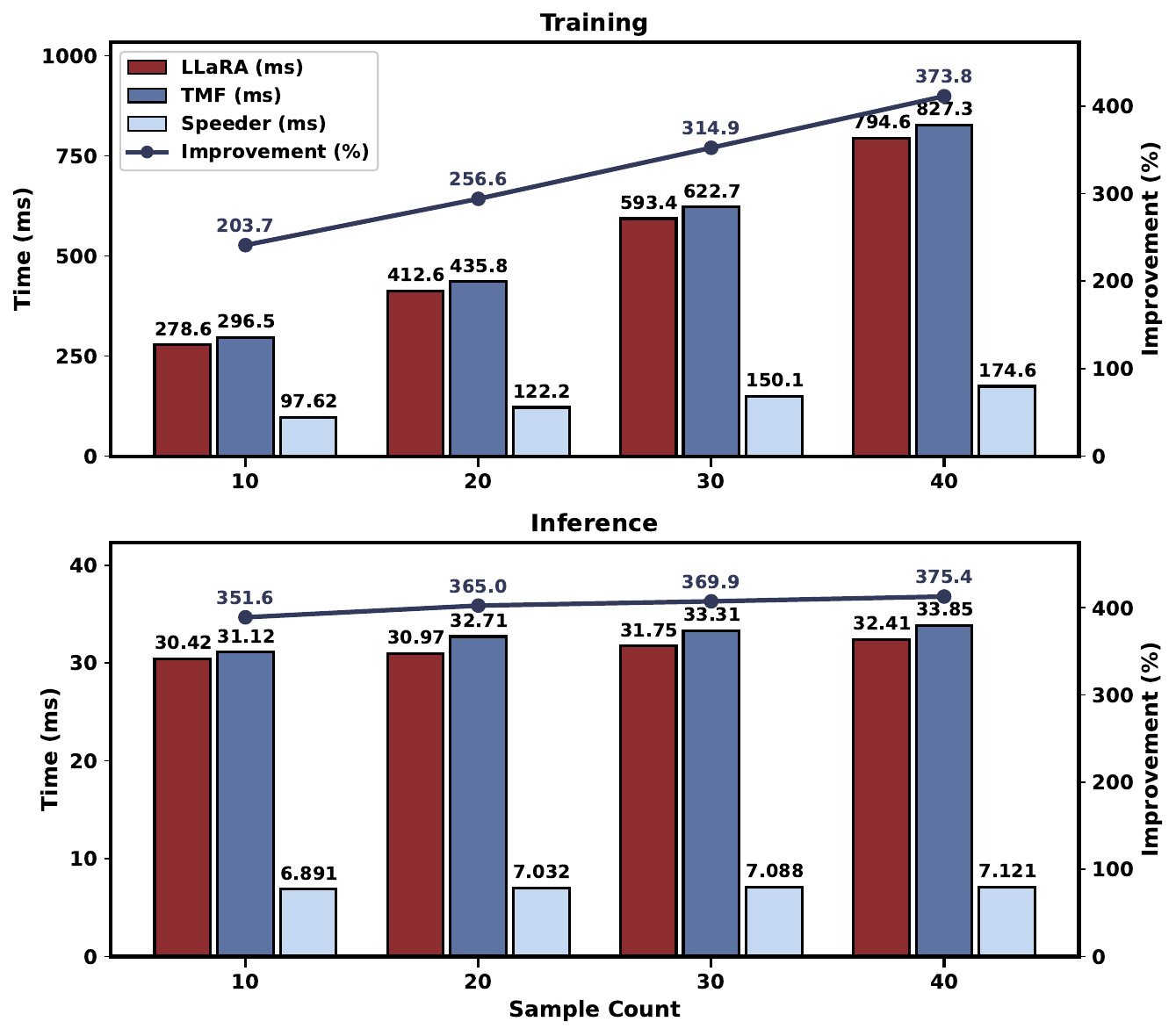}

\vspace{0.00 in}

\caption{Time consumption with different prompt lengths.} 
\label{improvement}
\vspace{0 in}
\end{figure}

We conducted an experiment by fixing the token length per item to 20 (so ($Avg_{token}$) is fixed to 20) and setting the candidate pool size $m$ to 5, 
in order to observe how the per-sample training and inference time consuming 
varies with different values of the number of interacted items $n$. The result is shown in~\Cref{improvement}.
The results demonstrate that as the number of interacted items $n$ grows during training, \ourmodel exhibits a substantially smaller increase in per-sample time consumption compared to LLaRA and TMF, indicating a more efficient and controllable scaling behavior.

In inference, after embedding caching, the time complexity is:
\begin{equation}
O\left(L \cdot t_{out} \cdot (N \cdot d + d^2)\right), \tag{Eq.4}
\end{equation}
here, $t_{out}$ denotes the length of the output context, and $N = Avg_{token} \cdot (n + m) + t_0$. $t_{out} \approx Avg_{token} \approx 20$ for LLaRA and TMF (as they output full titles), while $t_{out} = 1$ for Speeder (only index is returned). 
As displayed in~\Cref{improvement}, the time consuming of LLaRA, TMF and \ourmodel during inference rises moderately as $n$ grows, but the time consuming of \ourmodel remains the minimum, indicating its tremendous scalability during inference.

\nosection{The Evolution of Efficiency Improvement over \textbf{LLaRA} and \textbf{TMF} as Data Scale Increases} As the data scale increases, \ourmodel’s efficiency improvement over LLaRA and TMF continue to expand and eventually converge.
During training, substituting the expression $N = Avg_{token} \cdot (n + m) + t_0$ into Eq.3 with growing $n$ :
\begin{equation}
C_{train}^{{mo}}(n) = O(k_1^{mo} \cdot n^2 + k_2^{mo} \cdot n + k_3^{mo}), \tag{Eq.5}
\end{equation}
where $C_{train}^{mo}(n)$ stands for time complexity of model during training, mo is short for model, with constants $k_1 = (Avg_{token})^2 \cdot d$, $k_2 = Avg_{token} \cdot (2kd + d^2)$, $k_3 = kd(k + d)$, and $k = m \cdot Avg_{token} + t_0$.

The efficiency improvement over baseline model $m_1$ (LLaRA or TMF) is:
\begin{equation}
I_{train}(n) = \frac{C_{train}^{m_1}(n)}{C_{train}^{Speeder}(n)} - 1. \tag{Eq.6}
\end{equation}
Its first derivative with respect to $n$ is positive (proof omitted), meaning $I_{train}(n)$ is monotonically increasing with $n$.

As $n \to \infty$, the upper bound becomes:
\begin{equation}
\left(\frac{Avg_{token}^{LLaRA/TMF}}{Avg_{token}^{Speeder}}\right)^2 = \left(\frac{20}{2}\right)^2 - 1 = 99, \tag{Eq.7}
\end{equation}
implying a theoretical 99× speedup in training.

Similarly, it can be proved that $I_{inference}(n)$, \ie \ourmodel’s efficiency improvement during inference, is also monotonically increasing with respect to $n$. 
Likewise, based on Eq.4, it can be further derived the upper bound:
\begin{equation}
\left(\frac{t_{out}^{LLaRA/TMF}}{t_{out}^{Speeder}}\right) \cdot \left(\frac{Avg_{token}^{LLaRA/TMF}}{Avg_{token}^{Speeder}}\right) = \frac{20}{1} \cdot \frac{20}{2} - 1 = 199, \tag{Eq.8}
\end{equation}
indicating a 199× theoretical inference speedup.

In summary, when $n$ is small, the contribution of $t_0$ in Eq.3 and Eq.4 is non-negligible, thus reducing the observed efficiency gap. 
However, since $I_{{train}}(n)$ and $I_{{inference}}(n)$ inference analog are both monotonically increasing in $n$, \ourmodel’s advantage grows consistently with longer sequences and approaches the theoretical upper bound as $n \to \infty$.

As observed in~\Cref{improvement}, \ourmodel consistently outperforms LLaRA and TMF in both training and inference efficiency. The efficiency improvement exhibits progressively larger gains during training and maintains a consistently high advantage in inference.